\documentclass[preprint,12pt]{elsarticle}

\usepackage{amssymb}
\usepackage{amsmath}
\usepackage{color}
\usepackage{braket}
\usepackage[utf8]{inputenc}
\newcommand*{\Z}{\mathbb{Z}}

\journal{PhysicaD}

\begin{document}

\begin{frontmatter}

\title{Harmonic Structure of the Brunel spectra}

\author{Paul David, Fabrice Catoire, and Luc Berg{\'e}} 

\affiliation{organization={Université de Bordeaux-CNRS-CEA, Centre des Lasers Intenses et Applications (CELIA)}, city={Talence Cedex}, postcode={33405}, country={France}}

\begin{abstract}
Electromagnetic emissions, known as Brunel radiations, are produced in plasmas through the coupling between the free electron density and ultrafast ionizing laser pulses. The radiation spectrum generated in laser-gas interactions is here investigated from a local current model for laser drivers with two frequency components - or 'colors' - being not necessarily integers of one another. We provide a general description of this spectrum by deriving analytically the convolution product of the Fourier transforms of the electron density and of the laser electric field. Our analysis reveals that the only knowledge of the optical field extrema in time domain is sufficient to reproduce faithfully the numerically-computed Brunel spectrum and justify the emergence of various resonance frequencies. The classical combination of two laser harmonics, i.e., a fundamental and its second harmonic, is also addressed. 
\end{abstract}


\end{frontmatter}

\section{Introduction}
\label{sec1}

Brunel radiation \cite{Brunel1990} proceeds from the conversion mechanism by which a light wave generates secondary radiations from the temporal variations of the electron density current. For non-relativistic plasmas, this current density, ${\vec J}(t)$, is mainly driven by the product between the free electron density $N_e(t)$ and the pump laser electric field ${\vec E}(t)$ \cite{Berge2007}, i.e., 
\begin{equation}
    \frac{\partial {\vec J}}{\partial t} + \nu_c {\vec J} = \frac{e^2}{m_e} N_e(t){\vec E}(t),
    \label{1}
\end{equation}
where $e$, $m_e$ and $\nu_c$ denote the elementary charge, the electron mass and a collisional rate of free electrons with neutral atoms, respectively. For a single-frequency (or single-'color') pump field, it is well-known that the product $N_e(t) {\vec E}(t)$ is responsible for the creation of uneven harmonics of the carrier wave, as the electron density -- only depending on the length (absolute value) of the laser electric field -- decomposes in Fourier series over twice the pump frequency \cite{Brunel1990}. In contrast, for a two-color field containing, e.g., a fundamental laser frequency and its second harmonic (named the '$\omega-2\omega$ scheme' below), the Fourier spectrum of $N_e(t)$ comprises all harmonics of the fundamental and, therefrom, Brunel radiations are emitted at all integers of that pump frequency. This conversion mechanism explains, in particular, the generation of intense and ultrabroadband terahertz (THz) waves due to photocurrents induced along the early stage of plasma formation through the photoionization process \cite{Kim2008}. These photocurrents efficiently build up a low-frequency component when the temporal profile of the optical field is rendered asymmetric in time by adding a second color.

Although the $\omega-2\omega$ scheme for THz generation has been the topic of extensive research in the past years \cite{Wang2009,Liu2010,Babushkin2010,Babushkin2011,Gonzalez2015,Clerici2013,Nguyen2017,Nguyen2018_2,Damico2007,Andreeva2016,Nikolaeva2024,Vaicaitis2019,Babushkin2022,Kostin2019,Kostin2020,Vvedenskii2014}, much lesser attention has been paid to the generation of secondary radiations produced by nonharmonic pump frequencies \cite{Kostin2016,Vaicaitis2018,Kim2009,Wang2017,Zhang2017}. Such dedicated studies would, however, be all the more impactful that the available ultrafast optical parametric amplifiers do not generally deliver light waves operating at commensurable frequencies, while they make it possible to create secondary radiations over a broad frequency range, spanning from the mid-IR to ultraviolet \cite{Vaicaitis2018}. Indeed, whereas the $\omega-2\omega$ scheme only produces integers of the fundamental laser pump, including the ''$0$th order'' harmonic corresponding to broadband THz fields \cite{Babushkin2022}, mixing up pulses with incommensurate frequencies can supply a much richer spectrum \cite{Theberge2006,Theberge2010}. A related question is then to know whether varying the relative phase between two pump components with incommensurate frequency ratio may affect their conversion efficiency into the generated new radiation. For instance, a $\pi/2$ phase angle is expected to optimize photocurrent-induced terahertz pulse generation \cite{Kim2008,Li2012}. 

By numerically integrating Eq. (\ref{1}) for a linearly polarized laser electric field:
\begin{equation}
    {\vec E}(t) = {\vec E}_{\omega_1}(t) \cos{(\omega_1 t)} + {\vec E}_{\omega_2}(t) \cos{(\omega_2 t + \phi)},
    \label{2}
\end{equation}
where ${\vec E}_{\omega_1}(t)$ and ${\vec E}_{\omega_2}(t)$ are the amplitudes of the two laser colors at the respective frequencies $\omega_1$ and $\omega_2$, such that $\omega_i \neq 2 \omega_j$ ($i\neq j =1,2$), and $\phi$ being their relative phase, Kim \cite{Kim2009} reported the emission of Stokes $(2\omega_1 - \omega_2)$ and anti-Stokes $(2\omega_2 - \omega_1)$ satellites created via four-wave mixing (FWM) with the nonlinear electron current. The analytical treatment of the generation process proposed by the same author, however, cannot be applied to such a configuration, as it relies on an ionization rate driven by a single-color only, thus unable to reproduce the complete spectrum of the excited electron density. Besides experimental measurements \cite{Wang2017,Zhang2017}, THz pulse generation by the mutual enhancement of Brunel harmonics \cite{Kostin2019}, three-color mixing \cite{Wang2022} or when varying the frequency ratio $\omega_2/\omega_1$ \cite{Kostin2016} were also analytically addressed by Kostin and co-workers on the basis of Fourier series applied, e.g., to a tunneling-type ionization rate \cite{Ammosov1986}, either from a perturbation theory assuming a weak component $E_{\omega_2}$ or for multi-colored periodic pulses supposed maximum at instants maximizing each of the pulse components. Such assumptions do not hold for a general nonharmonic wave mixing, for which a non-periodic pump field does not admit any Fourier decomposition.

The current lack of straightforward analytical tools to describe  the generation of secondary Brunel radiations by gas plasmas - even out of the traditional $\omega-2\omega$ scheme - invites us to revisit the modeling of Brunel harmonic spectra within a systematic method. Here, we propose an original semi-analytical treatment of Brunel spectra allowing us to retrieve and understand the emergence of FWM spectral satellites for arbitrary frequencies $\omega_1$ and $\omega_2$, which are excited at the same resonant modes as those driven by the Kerr-driven response \cite{Agrawal2012} of the irradiated gas \cite{Theberge2006,Theberge2010,Vaicaitis2018}. Using a simple Taylor expansion of a multi-colored pump field near its extrema in time, $t_n$ -- being not necessarily zero or equal for all laser components -- we prove that the only knowledge of these time instants fixing the maxima in the ionization rate is enough to reproduce the overall Brunel spectrum. The central idea is to exploit the fact that the source term for the radiated fields described by Eq. (\ref{1}) is the convolution product of the spectra of the electron density and optical field \cite{Nguyen2018}, since, according to Jefimenko theory \cite{Jefimemko1989}, radiated electromagnetic emission in the far-field mainly follows from the time derivative of the current density, i.e., $E_{\rm rad}(t) \propto \partial_t {\vec J}$. The ionization rate causing the increase in the free electron density is here taken as the standard quasi-static tunneling (QST) ionization probability \cite{Ammosov1986,Rae1992,Landau1965,Gonzalez2014}, which reduces for a hydrogenoid atom gas and laser intensities larger than a few $10^{13}$ W/cm$^2$ to 
\begin{equation}
W\left[E(t)\right]=\frac{4 (U_i/U_H)^{\frac{5}{2}} \nu_a}{|E(t)/E_a|}\exp{\left(-\frac{2(U_i/U_H)^{\frac{3}{2}}}{3|E(t)/E_a|}\right)},
\label{3}
\end{equation}

with $U_i$ being the ionization energy of the gas, $\nu_a = 4.13 \times 10^{16}$ Hz, $E_a = 5.14 \times 10^{11}$~V/m and $U_H = 13.6$ eV is the ionization potential of hydrogen. Although the following analysis will be applied to H atoms, it would generally hold for more complex atomic systems. Equation (\ref{3}) governs the electron density growth described by the standard rate equation:
\begin{equation}
\frac{\partial N_e}{\partial t} = W(E(t))(N_a - N_e(t)),
\label{4}
\end{equation}
where $N_a$ is the density of neutral atoms.

The paper is organized as follows: Section \ref{sec2} specifies the vectorial laser field Eq. (\ref{1}) that will describe four typical polarization states of interest. The local current (LC) model constituted by the set of ordinary differential equations (\ref{1}) and (\ref{4}) employing the ionization rate Eq. (\ref{3}) and the input pulse Eq. (\ref{2}) will first be numerically integrated for these polarization geometries applied to a 800-nm, 35-fs fundamental pulse coupled to a fraction of 1300-nm second component, as delivered by current optical parametric amplifiers \cite{Vaicaitis2018}. To preliminarily validate the results of the LC model, the one-dimensional (1D) Time-Dependent Schr{\"o}dinger Equation (TDSE) will be integrated to extract an ab-initio (quantum) evaluation of the free electron current density supplied by the continuum-continuum transitions that take place along ionization of hydrogen by linearly polarized two-color pulses. The extracted electron density will be used to compute the source term $N_e(t)\vec{E}(t)$ yielding the Brunel spectra, which will be shown to agree with the full continuum-continuum component of the current, justifying thereby the applicability of the LC model. Then, in Section \ref{sec3}, an analytical treatment of the QST rate (\ref{3}) around the discrete set of the ionization instants $t_n$ designing the steplike increase in the electron density will be utilized to solve Eq. (\ref{1}) in Fourier domain and proved to restore accurately the numerically-computed spectra. The spectral emissions expected from nonharmonic pump components will be discussed from their analytical expression. In particular, we shall address the mid-IR and visible radiations emerging as Brunel-type harmonics at the FWM frequencies $2\omega_i-\omega_j$ ($i\neq j, i,j=1,2$). Section \ref{sec4} will revisit the classical $\omega-2\omega$ configuration for which a new expression of the Fourier series for the electron density will be derived. Section \ref{sec5} will summarize our main results.

\section{Results from the LC model}
\label{sec2}

\begin{figure}[ht!]
\centering \includegraphics[width=0.9\columnwidth]{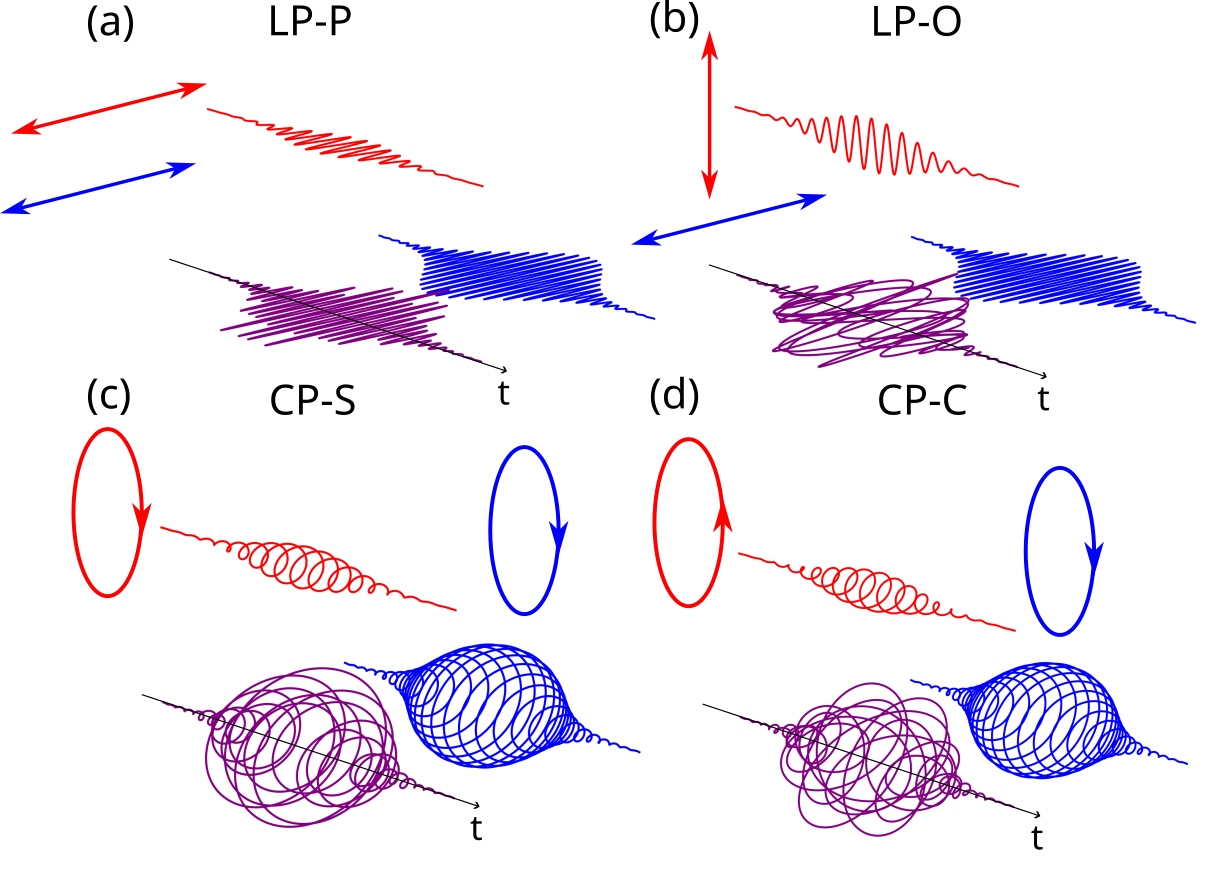}
\caption{Sketch of the four two-color baseline polarization states: (a) LP-P, (b) LP-O, (c) CP-S, and (d) CP-C for a fundamental pulse with frequency $\omega_1$ (blue pulse) coupled with a fraction of second color with frequency $\omega_2$ (red pulse, $\omega_2 = 0.615\,\omega_1$). The arrow represents the field polarization state of the different colors. Purple-colored distribution represents the sum of the time-dependent fields with an increasing time going to the right as indicated by the black arrow.}
\label{Fig1}
\end{figure}

To start with, we introduce the four polarization states of interest following Meng {\it et al.}'s convention originally employed in the context of the $\omega-2\omega$ scheme for two colors oriented in the $(x,y)$ plane \cite{Meng2016}, namely, i -- Linearly Polarized pulses with Parallel directions (LP-P), ii -- Linearly Polarized pulses with Orthogonal directions (LP-O), iii -- Circularly Polarized two-color pulses with Same helicity (CP-S), i.e, the two colors rotate in the same direction, and iv -- Circularly Polarized (CP) two-color pulses with counter-helicity (CP-C), i.e, the two colors rotate in opposite directions. These four polarization states are illustrated in Fig. \ref{Fig1}. They can be modeled using the vectorial Gaussian laser field \cite{Tailliez2020}: 
\begin{equation}
\label{5}
 \vec{E}(t)  =  \sum_{j=1,2}  \frac{E_{0,j} \mathrm{e}^{-2 \ln 2 \frac{t^2}{\tau^2_{j}}}}{\sqrt{1+\rho_j^2}}
 \left(\begin{array}{c}
 \cos(\omega_j t+\phi_j) \\
 \rho_j \cos(\omega_j t+\phi_j + \theta_j)
\end{array}\right) ,
\end{equation}
where the so-called ''fundamental'' (second) color with amplitude $E_1$ (resp. $E_2$) and central frequency $\omega_1$ (resp. $\omega_2$) is controlled by its respective ellipticity $\rho_j$, phase angle $\theta_j$, phase offset $\phi_j$ and Full-Width at Half-Maximum (FWHM) $\tau_j$ ($j=1,2$). The four baseline laser configurations then correspond to LP-P: $\rho_j=0$, CP-S: $\rho_j=1$, $\theta_1=\theta_2=\pm \pi/2$, CP-C: $\rho_j=1$, $\theta_1=-\theta_2=\pm\pi/2$ and LP-O: $\rho_1=0,\,\rho_2=+\infty$, $\theta_j=0$. Without loss of generality, we can reduce the phase angle to a simple relative phase between the two colors which we assume with same FHWM duration, i.e., $\phi_2 \equiv \phi,\,\phi_1 = 0$ and $\tau_{j=1,2} \equiv \tau$. The LP-P, CP-S and CP-C configurations can then be gathered as
\begin{equation}
\label{6}
{\vec E}(t) = \frac{E_0 \mathrm{e}^{-2 \ln 2 \frac{t^2}{\tau^2}}}{\sqrt{1+\rho^2}} \left [ \sqrt{1-r} \left(\begin{array}{c}
 \cos(\omega_1 t) \\
 \rho \cos(\omega_1 t + \theta)
\end{array}\!\right) 
+ \sqrt{r} \left(\begin{array}{c}
 \cos(\omega_2 t + \phi) \\
 \rho \epsilon \cos(\omega_2 t + \theta + \phi)
\end{array}\right) \right],
\end{equation}
restoring the LP-P, CP-S and CP-C pump cases by setting $\rho = 0$, $(\rho = 1,\,\epsilon = 1,\theta = -\pi/2)$ and $(\rho=1,\,\epsilon = -1,\theta = -\pi/2)$, respectively. The LP-O case is treated apart using
\begin{equation}
    {\vec E}_{\rm LP-O}(t) = E_0 \mathrm{e}^{-2 \ln 2 \frac{t^2}{\tau^2}} (\sqrt{1-r} \cos{(\omega_1 t)} {\vec e}_x + \sqrt{r} \cos{(\omega_2 t + \phi) {\vec e}_y}).
    \label{7}
\end{equation}

In Eqs. (\ref{6}) and (\ref{7}), $r$ denotes the intensity fraction of the second color, which is taken equal to $10\%$, i.e., $E_1 = E_0 \sqrt{1-r}$ and $E_2 = E_0 \sqrt{r}$ with $r = 0.0909$. In the following we shall consider a nonharmonic pulse mixing of the two central wavelengths $\lambda_1 = 0.8\,\mu$m ($\omega_1 = 2 \pi \times 375$ THz) and $\lambda_2 = 1.3\,\mu$m ($\omega_2 = 2 \pi \times 230.8$ THz), both associated with pulse envelopes of FWHM duration $\tau = 35$ fs. The fundamental component with intensity $I_1 \simeq 91$ TW/cm$^2$ ionizes the hydrogen gas, whereas the second color with intensity $I_2 \simeq 9.1$ TW/cm$^2$ does not contribute to liberate electrons. It is important to notice that, even if the second color alone does not ionize the medium, it structures the time evolution of the driving field and as such contributes to the spectrum of the electron density. We consider an electron–neutral collision rate of $\nu_c = 2.85$ ps$^{-1}$ and a neutral atom density of $N_a = 2.7\times 10^{19}$ cm$^{-3}$ \cite{Tailliez2020,Stathopulos2024}.\\

Before proceeding with the LC approach, we find it instructive to first validate the LC model by direct ab-initio computations in the specific case of a LP-P laser electric field. We thus performed simulations based on the Time-Dependent Schrödinger Equation (TDSE):

\begin{equation}
i\frac{\partial}{\partial t} \ket{\varphi(t)} = \left( \frac{(P+A(t))^2}{2} + V - \Phi(t) \right) \ket{\varphi(t)}.
\label{Eq:TDSE}
\end{equation}

\noindent In this equation formulated in atomic unit (a.u.), $P$ is the momentum operator, $V$ is the potential operator describing the electron-ion interaction from a soft-core potential so that the ionization potential of the ground state is $-0.5$ a.u. as for hydrogen atom. $A(t)$ and $\Phi(t)$ are the operators related to the vector and scalar laser potentials driving the laser-matter interaction. The TDSE is solved in 1D using a Discretized Variable Representation on a 1500 sequence of points. The box size is $2000$ a.u., allowing $80$ bound states to be included. To propagate the TDSE we used the first order of the Magnus expansion of the evolution operator and the Lanczos algorithm. A time step of $3\times10^{-2}$ a.u. is chosen and the Lanczos basis has been fixed to $50$ in order to ensure convergence of the calculation. The TDSE is solved in the length gauge, i.e., $A(t) = 0$ and $\Phi(t) = -zE(t)$. From the computation of $\ket{\varphi(t)}$ one can express the current associated with the emitted radiation. This current is explicitly written as $j = \bra{\varphi(t)} \Pi(t) \ket{\varphi(t)}$, where $\Pi(t) = P + A(t)$ is the generalized momentum, and contains all frequency components from THz to high-order harmonics \cite{Catoire2016}. The source term for secondary radiations follows from the time derivative $\partial_t {\vec j}(t)$ which can be computed using the Ehrenfest theorem (see, e.g. Ref. \cite{Vabek2022}). In this framework, the wave function $\varphi(t)$ is split into the sum of bound [$\varphi_b(t)$] and continuum components [$\varphi_c(t)$] \cite{Vabek2022}, in such a way that isolating the contribution associated with continuum-continuum transitions of the full current yields $\partial_t {\bf j}_{\rm cc} = -N_a\bra{\varphi_c(t)} \nabla (V -\phi) \ket{\varphi_c(t)}$. Note that we have introduced the gas density $N_a$ to link the microscopic current obtained from the TDSE to the macroscopic current used in Maxwell equations \cite{Finke2022}. From the TDSE we can also extract the electron density $N_e^{\rm TDSE}(t)$ in order to compute the quantum equivalent of the LC-model source term, namely $N_e^{\rm TDSE}(t)E(t)$ \cite{Vabek2022}.

\begin{figure}[ht!]
\centering \includegraphics[width=\columnwidth]{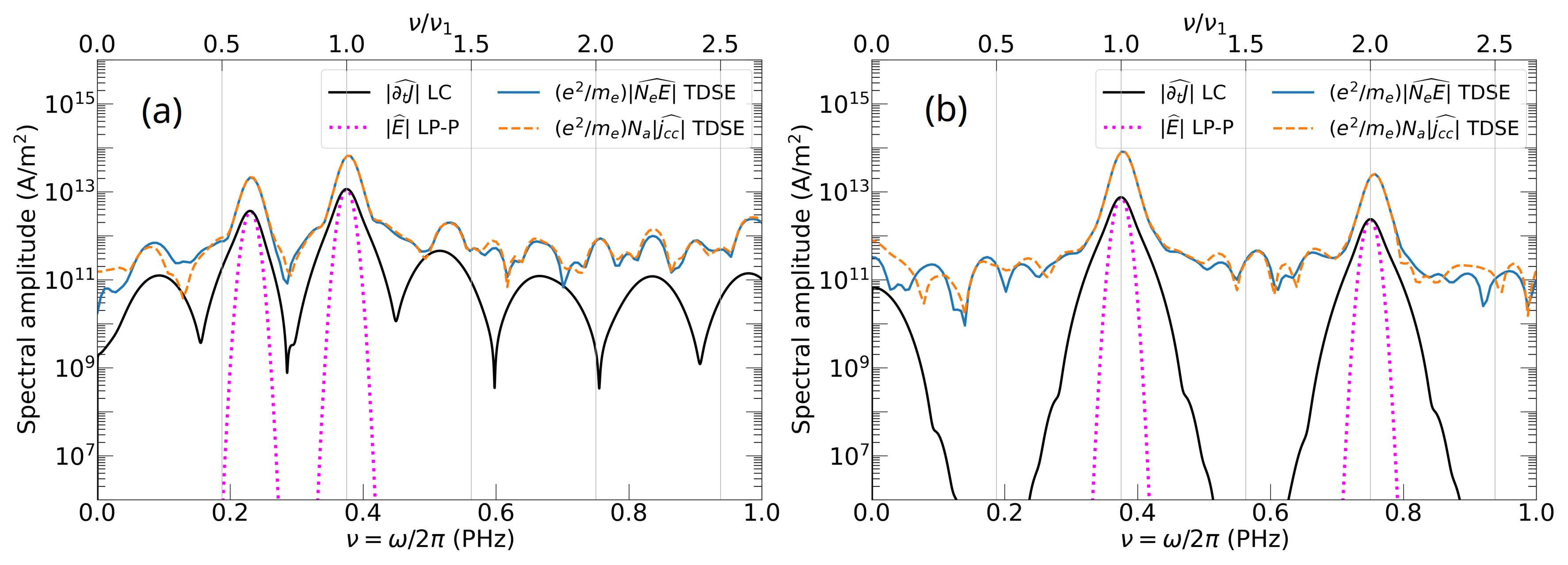}
\caption{Spectra from TDSE ab-initio simulations vs. LC spectra for 35-fs LP-P Gaussian pulses ionizing hydrogen. The current obtained from continuum-continuum transitions is plotted as a blue solid curve. The Fourier transform of $Ne(t)E(t)$ where $N_e(t)$ is extracted from the TDSE computations is plotted as an orange dashed curve. The LC spectra are plotted as black solid curves: (a) for a nonharmonic frequency mixing with frequency $\omega_2 = 0.615\,\omega_1$ and (b) in the $\omega - 2\omega$ scheme with $\omega_2 = 2 \omega_1$. The fundamental wavelength is 0.8 $\mu$m. The LP-P pump field spectrum $|\widehat{E}|$ is shown as pink dotted lines.}
\label{Fig2}
\end{figure}

The results are presented in Fig. \ref{Fig2}, both for a nonharmonic pulse mixing [Fig. \ref{Fig2}(a)] and for the $\omega - 2\omega$ scheme [Fig. \ref{Fig2}(b)]. The first important observation is the similar spectral distributions obtained from the quantum electron current $\partial_t {\bf j}_{\rm cc}$ and the one computed from $N_e^{\rm TDSE}(t) E_L(t)$. This confirms, from a quantum perspective, the relationship $\partial_{t} j_{cc} \simeq N_e^{\rm TDSE}(t) E(t)$ that is currently exploited in the LC model. We also plotted the LC spectra for comparison. We can notice that, for incommensurate pump frequencies [Fig. \ref{Fig2}(a)], all the main spectral peaks expected from the LC spectra are well reproduced by the TDSE. Extra peaks, however, appear in the region $1.7 \,\nu_1 - 2.2 \,\nu_1$ of the TDSE spectra, suggesting an undergoing complex dynamics, the explanation of which is beyond the scope of the present paper. The amplitude of the LC spectra is also lower compared to the TDSE results. This particular discrepancy is due to the over-simplification of the QST rate used in Eq. (\ref{3}). A more precise ionization rate could be employed to compare better with the TDSE data, but it would only affect the quantitative values of the spectral amplitude and not the dynamics.

We have also performed TDSE calculations for the $\omega-2\omega$ scheme in the LP-P configuration. The results are provided in Fig. \ref{Fig2}(b). Again, the TDSE results obtained from the computation of the continuum-continuum contribution of the current is in good agreement with the source term $N_e(t)E(t)$ of the Brunel's emission mechanism. We have also plotted the result obtained from the LC model. The main contributions appear at the correct pump frequencies $\nu_1 , 2 \nu_1$ and in the THz range. Again, as in the incommensurable case, a discrepancy occurs in the amplitude, attributed to the simplified formulation of the ionization rate. Additional contributions in between the main spectral peaks are inherent to the complex continuum-continuum dynamics along the ionization process, which cannot be described by a simple rate equation (see for instance \cite{Vabek2022}). Having validated the LC approach from ab-initio simulations, we detail below the key physical players building up the structure of Brunel spectra.

\begin{figure}[ht!]
\centering \includegraphics[width= \columnwidth]{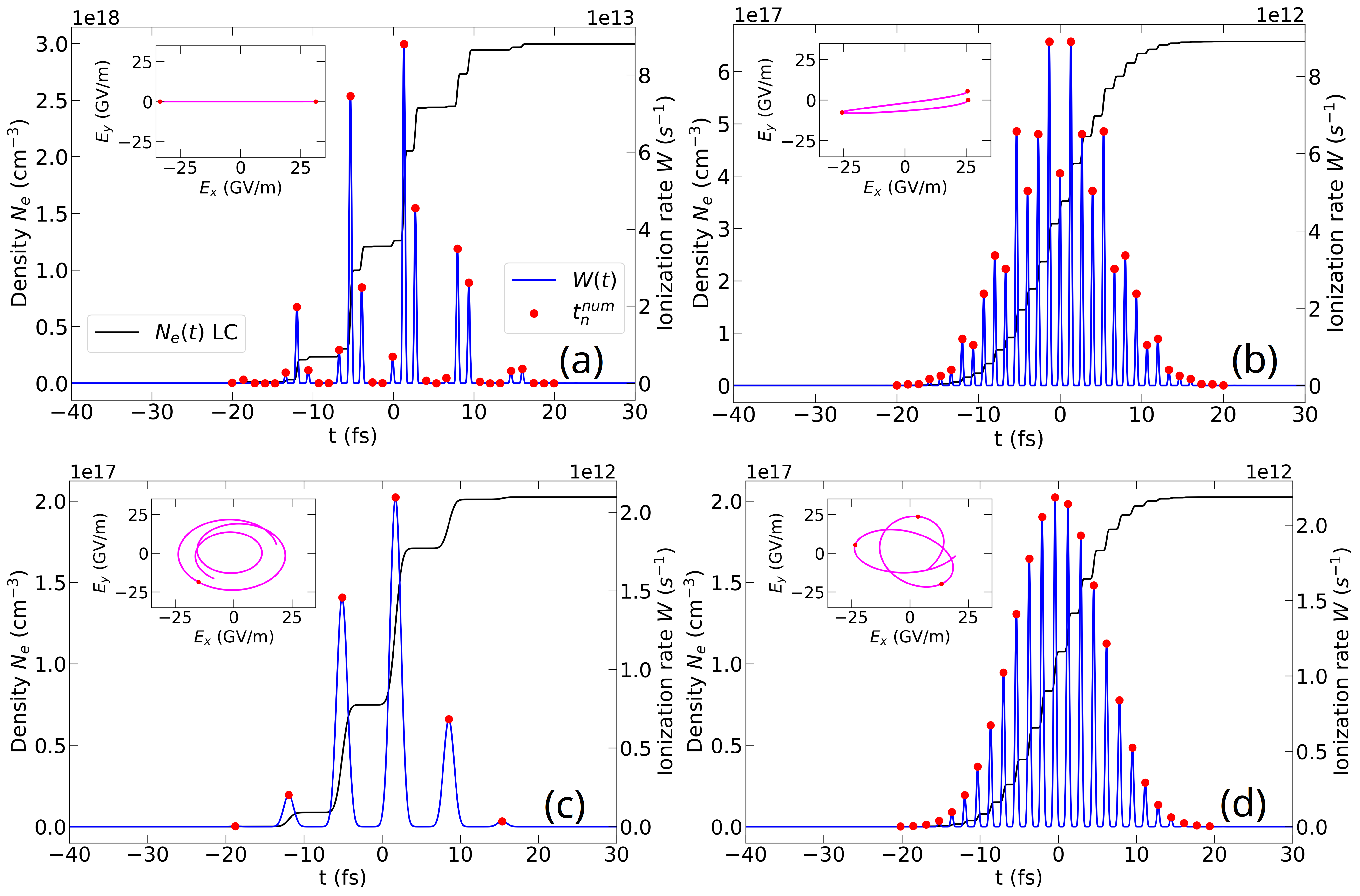} 
\caption{(a-d) Density profiles (solid black curves, left axis), ionization rates (solid blue curves, right axis) and field polarization patterns in the $(E_x,E_y)$ plane (insets) of 35-fs, two-color Gaussian pulses for the four polarization geometries of interest (see text for the laser-gas parameters). The two-color relative phase is $\phi = \pi/2$. The Lissajou figures shown as insets run over different time intervals close to their characteristic periods and capturing the highest extrema of $W(t)$, namely, LP-P and LP-O: $0-2.72$ fs, CP-S: $0-7$ fs, CP-C: $-3 - 2.5$ fs. The (exact or analytical) ionization times $t_n$ are plotted in the time window from - 21 fs to + 21 fs.}
\label{Fig3}
\end{figure}

Figure \ref{Fig3} illustrates the field patterns, ionization yields and QST rates for the above two-color 35-fs Gaussian pulses ionizing hydrogen in their LP-P, CP-S, CP-C and LP-O configurations. The relative phase is chosen as $\phi = \pi/2$, which is known to optimize THz field generation in the LP-P $\omega-2\omega$ scheme. The first observation is the loss of periodicity in the ionization responses under the pulse envelope. The highest density levels are reached for the LP-P pulse inducing two ionization events per cycle lined up on the $\omega_1$ frequency (with optical cycle $\tau_{\rm o.c.}^{\rm LP} \simeq 2.7$ fs). The electron density of an LP-O pulse is smaller by a factor $\sim 4.5$. By comparison, the $1/\sqrt2$ factor affecting the amplitude of the circularly polarized electric fields induces a consecutive weakening in $N_e(t)$ by about one order of magnitude [see bottom Figs. \ref{Fig3}(c,d)]. Ionization develops over longer and $2 \pi/(\omega_2 - \omega_1)$-periodic ionization steps (with corresponding optical cycle $\tau_{\rm o.c.}^{\rm CP-S} \simeq 6.93$ fs). Remarkably, the CP-C pulse develops three times more ionization events being regularly spaced over an ultrashort $2 \pi/(\omega_1 + \omega_2)$ periodicity (resp., $\tau_{\rm o.c.}^{\rm CP-C} \simeq 1.65$ fs). The electric field vector trajectories (Lissajou figures) in the $(E_x,E_y)$ plane have also been plotted within time intervals close to their respective optical cycles and surrounding maximum values of $W(t)$ (see insets). These patterns reveal two (three) main ionizing extrema in the LP-P (resp. LP-O) pump amplitude, which contribute to the ionization yield over the time window $0-2.72$ fs. Note the flattened contours of the CP-S pulse due to the non-zero $\omega_2$ component and the almost perfect triangular distribution of CP-C pulses at their field maxima where ionization takes place. In the CP-S case, the presence of a unique ionization event is expected to provide maximum conversion efficiency for equal electron density into the mid-IR/THz frequency interval \cite{Meng2016}. In the CP-C case, the triangular location of the three field maxima with sub-2 fs periodicity can make the same radiation yield vanish for $>10$ fs long driving pulses \cite{Tailliez2020}.

\begin{figure}[ht]
\centering \includegraphics[width=\columnwidth]{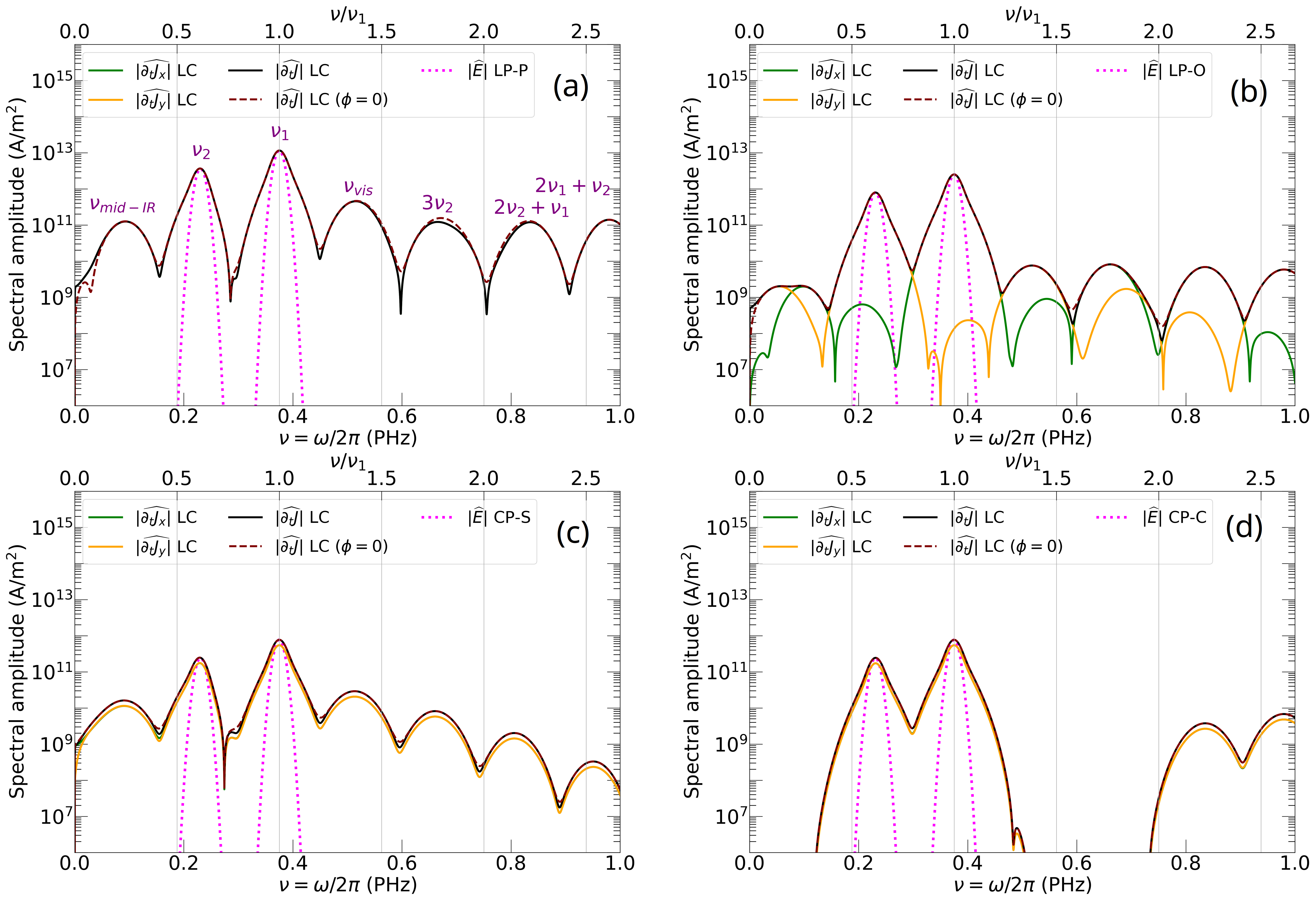}
\caption{Brunel radiation spectra numerically-computed from the LC model for the four polarization configurations (a) LP-P, (b) LP-O, (c) CP-S, and (d) CP-C as functions of $\nu = \omega/2 \pi$. The spectral amplitudes of the current components $\widehat{\partial_t J_x}$ and $\widehat{\partial_t J_y}$, of the total induced current density $|\widehat{\partial_t J}| = (|\widehat{\partial_t J_x}|^2 + |\widehat{\partial_t J_y}|^2)^{1/2}$ (see legend), and of the input pump with maximum lined up on max$\widehat{|\partial_t J}|$ for $\phi = \pi/2$ (pink dotted lines) are shown. Brown dashed curves illustrate $|\widehat{\partial_t J}|$ for a zero relative phase.} 
\label{Fig4}
\end{figure}

Following the LC model, the above waveforms, ionization rate [provided by Eq. (\ref{3})] and electron density [computed by Eq. (\ref{4})] are next directly plugged into Eq.~(\ref{1}). Figure \ref{Fig4} details the radiated spectra inferred from the Fourier transform of the resulting time-derivative of the electron current -- hat symbol refers to the Fourier transform in time using the convention $\widehat{f}(\omega) = \int f(t) \exp{(i \omega t)} dt/\sqrt{2\pi}$. Besides the input spectra centered at $\nu_1 = 375$ THz and $\nu_2 \simeq 231$ THz (dotted curves), we can observe the Stokes and anti-Stokes FWM components located at $\nu_{\rm vis} = 2 \nu_1 - \nu_2 = 519$ THz and $\nu_{\rm mid-IR} = 2 \nu_2 - \nu_1 = 87$ THz, respectively, together with higher-order spectral satellites generated by either third-harmonic generation or cascading process mixing up integers of the fundamental and second color [$3\nu_2,\,2\nu_2+\nu_1,\,2\nu_1+\nu_2$ and $3\nu_1$ in increasing order - see Fig. \ref{Fig4}(a)]. These new resonant modes are those similarly triggered by a Kerr-type nonlinear polarization ${\vec P}_K = \epsilon_0 \chi^{(3)} {\vec E}^3$, $\epsilon_0$ and $\chi^{(3)}$ denoting the permittivity in vacuum and the third-order electric susceptibility, respectively. In the region $\nu \rightarrow 0$ we can observe that, as predicted in Refs. \cite{Nguyen2018,Thiele2017}, an LP-O pump arrangement produces a THz yield which is conveyed by the second color and reaches a smaller amplitude (here by a factor $\sim 1/5$) than that of its LP-P counterpart [compare Figs. \ref{Fig4}(a) and \ref{Fig4}(b)]. By contrast, the most intense, generated mid-IR radiation ($\nu = 2\nu_2 - \nu_1$) is mainly carried out by the fundamental pulse (along ${\vec x}$). The visible radiation ($\nu = 2\nu_2 - \nu_1$) is conveyed by the second color (along ${\vec y}$). All pulse configurations are barely sensitive to switching the relative phase $\phi$ from $\pi/2$ to zero, in agreement with \cite{Kim2009} and in contrast with the $\omega - 2\omega$ scheme addressed in Sec. \ref{sec4}. We can infer a higher THz amplitude reached by a CP-S laser field compared with a LP-P one at equal ionization yield, when one keeps in mind the tenfold decrease of the electron density own to circular polarization [see Fig. \ref{Fig3}(c)]. Finally, we report the quasi-extinction of any secondary radiation when utilizing a CP-C pulse. First revealed in the THz frequency range \cite{Meng2016,Tailliez2020}, this property now seems to manifest both in the mid-IR and the visible range and can be attributed to the highly symmetric distribution of the ionization events giving birth to electromagnetic radiation that destructively interferes over a few optical cycles of the fundamental pump.

\section{Theoretical treatment}
\label{sec3}

Our theoretical method proceeds in the following three steps:

\begin{itemize}
    \item First, we approximate the steplike temporal dynamics of the free electron density according to the local photocurrent theory of Ref. \cite{Babushkin2011}. Ionization events
occur in the neighborhood of the electric field extrema located at $t_n$ such that $|{\vec E}(t)| \approx |{\vec E}(t_n)| + \frac{1}{2}\partial_t^2 |{\vec E}|(t_n)(t-t_n)^2$ and the QST ionization rate (\ref{3}) results from the superposition principle summing up all discrete ionization contributions around $t=t_n$, i.e., 
\begin{equation}
    \label{8}
    W(t) \approx \sum_n W(|{\vec E}(t_n)|) \mbox{e}^{-(t-t_n)^2/\tau_n^2},
\end{equation}
where
\begin{equation}
    \label{9}
    \tau_n = |E(t_n)|\sqrt{3 (U_H/U_i)^{3/2}/(E_a |\partial_t^2 E(t_n)|)}
\end{equation}
represents the duration of the $n$th ionization event. 
The electron density can thus be approximated as
\begin{equation}
    \label{9b}
N_e(t) \approx \sum_n \delta N_n H_n(t-t_n);\,\,\,H_n(t)=\frac{1}{2}[1+\mbox{Erf}(\frac{t}{\tau_n})],
\end{equation}
where $H_n(t)$ are steplike functions increasing at each instant $t_n$ and $\delta N_n$ is the $n$th density jump. The sum of the density jumps over all the ionization events determines the degree of ionization. With $N_e(t)=N_a[1-\exp(-\int_{-\infty}^t W(\tau) \,d\tau)]$ and $\int_{-\infty}^t W(\tau) \,d\tau \approx
\sum_n \sqrt{\pi} \tau_n W(t_n) H_n(t-t_n)$ it is possible to establish by recurrence the explicit form of $\delta N_n$ \cite{Gonzalez2015}:
\begin{align}
\label{10}
\delta N_1 & = N_a \left( 1- {\rm e}^{-\sqrt{\pi}\tau_1 W(t_1)} \right), \\
\delta N_n & = N_a {\rm e}^{-\sum_{m=1}^{n-1} \sqrt{\pi}\tau_m W(t_m)  } \left( 1- {\rm e}^{-\sqrt{\pi}\tau_n W(t_n)} \right), \qquad n>1,
\label{10b}
\end{align}
accounting for the saturation of the ionization yield at high enough pump intensity. The extrema instants $t_n$ are given by the roots of the time derivative of $|{\vec E}(t)|$ [see Eq. (\ref{5})]. Either these roots are computed exactly by direct numerical calculation - including the derivatives of the pulse envelopes - or they can be approximated under certain assumptions. Current hypotheses consist in assuming that the temporal pulse envelopes are slowly-varying ($\omega_j \tau_j \ll 1$) and the second color amplitude is small compared with the fundamental pulse ($r \ll 1$) \cite{Babushkin2011,Tailliez2020}. Under these assumptions the ionization instants $t_n$ for $n \in \Z$  express as:
\begin{align}
    \label{11a}
    \mbox{LP-P}: & \quad \omega_1 t_n \approx n \pi - (-1)^n \sqrt{\frac{r}{1-r}}\frac{\omega_2}{\omega_1}  \sin{(\frac{\omega_2}{\omega_1} n \pi + \phi)}, \\ 
    \label{11b}
    \mbox{LP-O}: & \quad \omega_1 t_n \approx n \pi - \frac{1}{2} \frac{r}{1-r}\frac{\omega_2}{\omega_1} \sin{(2\frac{\omega_2}{\omega_1} n \pi + 2\phi)}, \\ 
    \label{11c}
    \mbox{CP-S}: & \quad t_n = \frac{2 n \pi - \phi}{\omega_2 - \omega_1}, \\
    \label{11d}
    \mbox{CP-C}: & \quad t_n = \frac{2 n \pi - \phi}{\omega_2 +\omega_1}. 
\end{align}
When ignoring the temporal variations of the envelope, Eqs. (\ref{11c}) and (\ref{11d}) are exact roots of $\partial_t |{\vec E}(t)| = 0$ in the case of circularly polarized pulses.

\item Second, we compute the spectrum of the free electron density given by the Fourier transform in time of $N_e(t)$, i.e.,
\begin{equation}
    \label{12}
    \widehat{N_e}(\omega) = \frac{i}{\sqrt{2 \pi}} \sum_n \delta N_n \frac{\mbox{e}^{i \omega t_n - \omega^2 \tau_n^2/4}}{\omega + i\kappa},
\end{equation}
which encompasses the ultrabroadband Gaussian spectrum of each attosecond ($\tau_n)$-scaled ionization event and the coherent superposition of each spectral contribution in $\mbox{e}^{i \omega t_n}$. The parameter $\kappa$ appearing in the denominator is used to regularize the electron density distribution in Fourier space, as the latter does not vanish once the driving field is over. $\kappa$ is typically of the order of the driving pulse duration.  

\item Third, Eq. (\ref{12}) is convolved with the Fourier transform of the vector pump field [provided by Eq. (\ref{5})], yielding the radiated spectrum
\begin{equation}
    \label{13}
    {\vec E}_{\rm rad}(\omega) = g (\frac{\omega}{\omega + i \nu_c}) \frac{e^2}{\sqrt{2\pi} m_e} \widehat{N_e} \ast {\widehat{\vec E}} (\omega),
\end{equation}
with $g$ being the Jefimenko factor \cite{Babushkin2011}.

For the Gaussian pulses (\ref{5}), the above convolution product expresses as
\begin{align}
\label{14}
    {\vec E}_{\rm rad}(\omega) & = (\frac{\omega}{\omega + i \nu_c}) \frac{i g e^2}{8\pi \sqrt{\ln{2}} m_e} \sum_n \delta N_n \sum_{j=1}^2 \frac{E_{0j} \tau_j}{\sqrt{1+\rho_j^2}} \\ 
    & \times \int_{-\infty}^{+\infty} \frac{\mbox{e}^{i (\omega -u) t_n - (\omega - u)^2 \frac{\tau_n^2}{4}}}{\omega - u + i\kappa}
    \left(\begin{array}{c}
 \sum_{\pm} \mbox{e}^{-\frac{\tau_j^2}{8 \ln2}(\omega_j \pm u)^2 \pm i \phi_j} \\
 \rho_j \sum_{\pm} \mbox{e}^{-\frac{\tau_j^2}{8 \ln2}(\omega_j \pm u)^2 \pm i (\phi_j+\theta_j)} \nonumber
\end{array}\right) du,
\end{align}
and has to be computed numerically. This expression can, however, be simplified in the slowly-varying approximation $\omega_j \tau_j \rightarrow +\infty$, from which contributions in $\tau_j \mbox{exp}{[-\tau_j^2(\omega_j \pm u)^2)/(8 \ln2)]}$ reduces to $2 \sqrt{2 \pi \ln{2}} \delta(\omega_j \pm u)$ where $\delta(x)$ is the usual Dirac delta function. We then obtain 
\begin{align}
    \label{14b}
    {\vec E}_{\rm rad}(\omega) & = (\frac{\omega}{\omega + i \nu_c}) \frac{i g e^2}{\sqrt{2\pi} m_e} \sum_n \delta N_n \sum_{j=1}^2 \frac{E_{0,j}}{\sqrt{1+\rho_j^2}}  \frac{\mbox{e}^{i \omega t_n - (\omega^2 + \omega_j^2) \frac{\tau_n^2}{4}}}{((\omega + i\kappa)^2 - \omega_j^2)} \times \\ \nonumber
    & \hspace{-2cm} \left(\begin{array}{c}
 \omega \cos{(\omega_j t_n + \phi_j + 2 i \omega_j \omega \frac{\tau_n^2}{4})} - i \omega_j \sin{(\omega_j t_n + \phi_j + 2 i \omega_j \omega \frac{\tau_n^2}{4})} \\
 \rho_j \omega \cos{(\omega_j t_n + \phi_j + \theta_j + 2 i \omega_j \omega \frac{\tau_n^2}{4})} - i \rho_j \omega_j \sin{(\omega_j t_n + \phi_j + \theta_j + 2 i \omega_j \omega \frac{\tau_n^2}{4})}
\end{array}\right),
\end{align}
from which the ionization contributions in $\cosh,\sinh{(2 \omega \omega_j \tau_n^2/4)}$ distort the input spectra. Since the ionization duration is typically of the order of $\omega_1 \tau_n \sim (|{\vec E}(t_n)|/E_a)^{1/2} \ll 1$ under our current assumptions, Eq. (\ref{14b}) may be simplified to
\begin{align}
    \label{15}
    {\vec E}_{\rm rad}(\omega) & \simeq (\frac{\omega}{\omega + i \nu_c}) \frac{i g e^2}{\sqrt{2\pi} m_e} \sum_n \delta N_n \sum_{j=1}^2 \frac{E_{0j}}{\sqrt{1+\rho_j^2}}  \frac{\mbox{e}^{i \omega t_n - (\omega^2 + \omega_j^2) \frac{\tau_n^2}{4}}}{((\omega + i\kappa)^2 - \omega_j^2)} \times \\ \nonumber
    & \left(\begin{array}{c}
 \omega \cos{(\omega_j t_n + \phi_j)} - i \omega_j \sin{(\omega_j t_n + \phi_j)} \\
 \rho_j \omega \cos{(\omega_j t_n + \phi_j + \theta_j)} - i \rho_j \omega_j \sin{(\omega_j t_n + \phi_j + \theta_j)}
\end{array}\right).
\end{align}

\end{itemize}

The Brunel spectra can thus be either provided by Eq. (\ref{13}) using the numerically-computed (or so-called ''exact'') instants $t_n^{\rm num}$ or approximated by Eq. (\ref{15}) using the same exact instants $t_n^{\rm num}$ or their estimates $t_n^{\rm an}$ given by Eqs. (\ref{11a})-(\ref{11d}) (valid for small amplitude ratios $r \ll 1$ for linearly polarized pulses). Checking the validity of Eq. (\ref{15}) with the ionization instants Eq. (\ref{11a})-(\ref{11d}) will allow us to understand the underlying physics of Brunel secondary emissions. 

\begin{figure}[ht!]
\centering \includegraphics[width=\columnwidth]{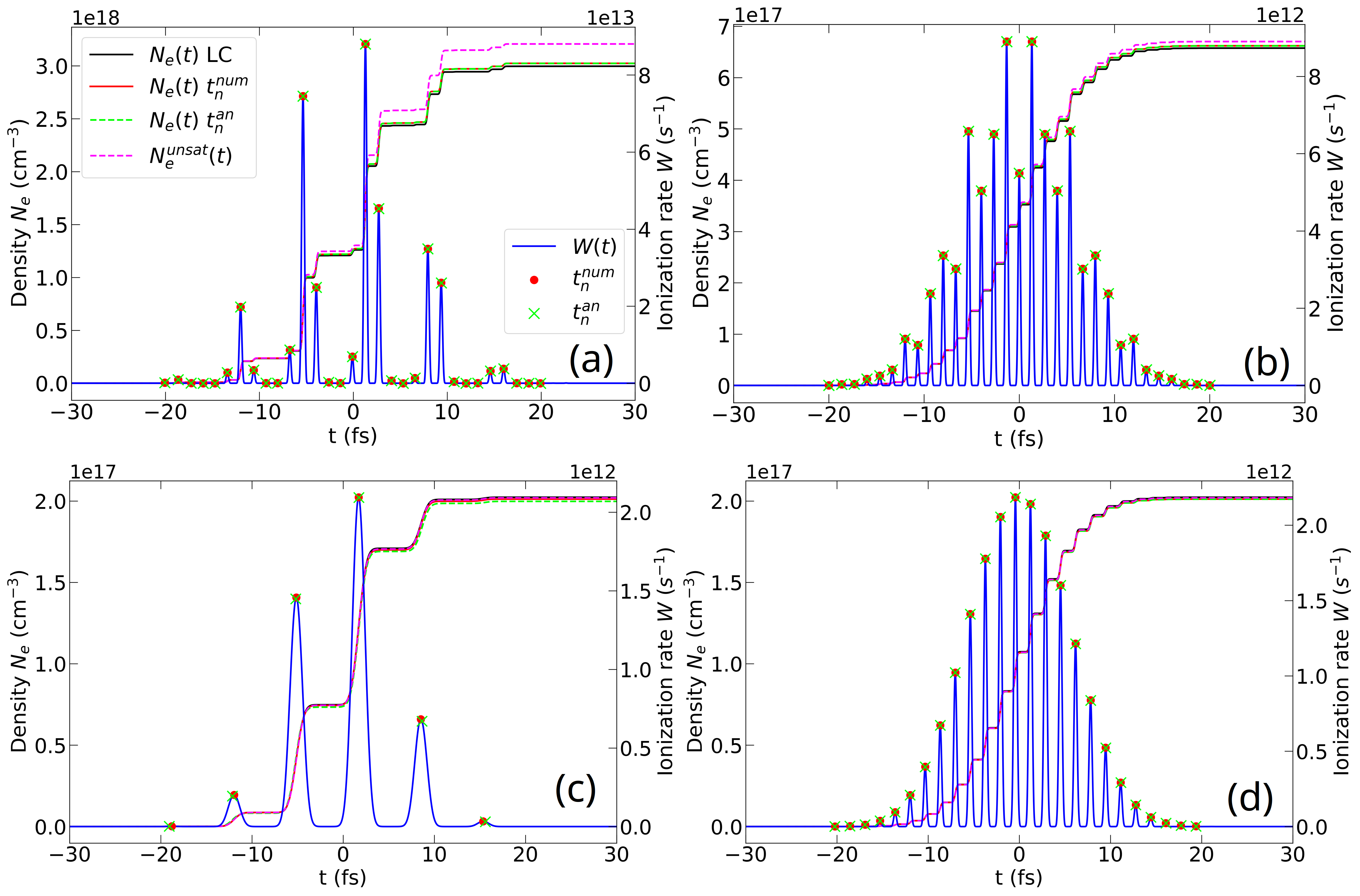}
\caption{Profile of the density growth (left axis) evaluated from Eq. (\ref{9b}) (red solid curve) using Eqs. (\ref{8})-(\ref{10b}) and the numerically-computed instants $t_n^{\rm num}$ for the four baseline polarization configurations (a) LP-P, (b) LP-O, (c) CP-S, and (d) CP-C and compared with their counterparts using the analytical instants $t_n^{\rm an}$ given by Eqs. (\ref{11a})-(\ref{11d}) (green dashed curves). The black solid curve recalls the exact numerically-computed density profiles from the LC model (Fig. \ref{Fig3}). The pink dotted curve show the unsaturated electron density using the simplified density step $\delta N_n = N_a \sqrt{\pi} \tau_n W(t_n)$. The ionization rates are also plotted as blue solid lines (right axis) and display the exact and analytical maximum instants $t_n^{\rm num}$ and $t_n^{\rm an}$ marked as red dots and green crosses, respectively. The (exact or analytical) ionization times $t_n$ are plotted in the time window from - 21 fs to + 21 fs.} 
\label{Fig5}
\end{figure}

To this aim, Fig. \ref{Fig5} displays, for the four baseline polarization states, the electron density profiles computed from Eq. (\ref{9b}) using the exact instants $t_n^{\rm num}$ (solid curves) or their estimates given by Eqs. (\ref{11a})-(\ref{11d}) ($t_n^{\rm an}$, dashed curves). For comparison, the numerically-computed densities extracted from Fig. \ref{Fig3} are recalled. The ionization steps $\delta N_n$ are computed from Eqs. (\ref{10}) and (\ref{10b}) using the value of $W(t_n)$ and the ionization duration (\ref{9}) at each of these instants. Remarkably, the resulting density profiles appear in excellent agreement with their LC counterpart, when employing both the exact $t_n$ and their analytical evaluations (\ref{11a})-(\ref{11d}). Possible tiny deviations can be attributed to our Taylor expansion of the laser field truncated to second order only. Note also the relative importance of the saturation effect missed when the approximation
\begin{equation}
    \label{15b}
N_e(t) \approx N_a \int_{-\infty}^{t} W(\tau)d\tau \equiv N_e^{\rm unsat}(t)
\end{equation}
is applied, which can severely alters the quantitative value of the electron density plateau reached asymptotically, e.g., in the LP-P case.

\begin{figure}[ht!]
\centering \includegraphics[width=\columnwidth]{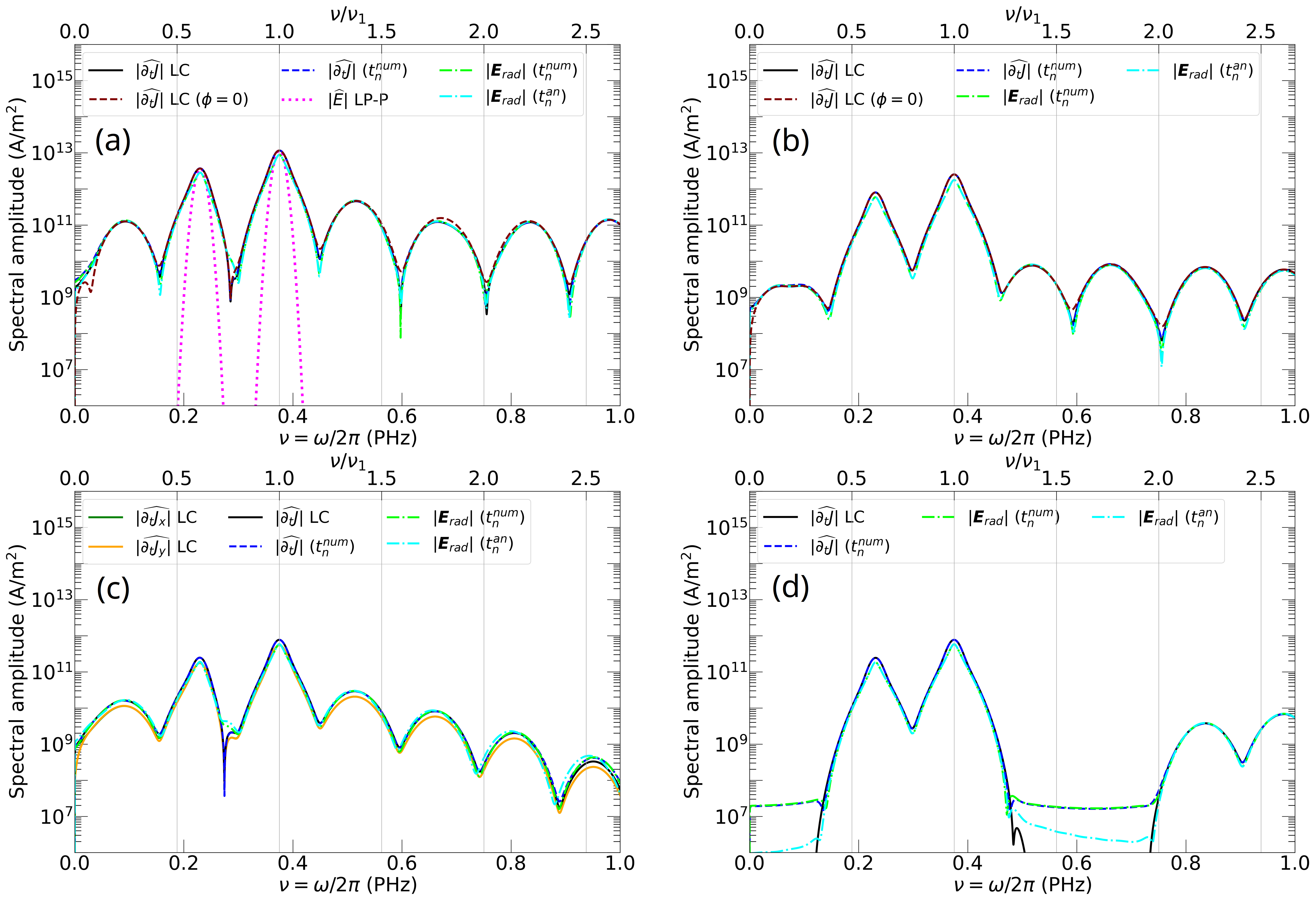}
\caption{Brunel spectra $|\widehat{\partial_t J}|$ computed from Eq. (\ref{13}) for the four polarization configurations (a) LP-P, (b) LP-O, (c) CP-S, and (d) CP-C as functions of $\nu = \omega/2 \pi$ for $\phi = \pi/2$ (black solid curves) or $\phi = 0$ (brown dashed curves), and compared to their approximations described by Eqs. (\ref{12}) and (\ref{13}) using the exact instants $t_n^{\rm num}$, and by Eq. (\ref{15}) using either $t_n^{\rm num}$ or their analytical evaluation $t_n^{\rm an}$ [Eqs. (\ref{11a})-(\ref{11d})] (see legend). The CP-S spectra evaluated with $|{\vec E}_{\rm rad}|$ are plotted from Eq. (\ref{14b}). The input pump has its maximum lined up on max$\widehat{|\partial_t J}|$ for $\phi = \pi/2$ (pink dotted lines, LP-P case only). Electron density spectra plotted from Eq. (\ref{12}) have been regularized with $\kappa = 5 \times 10^{-2}$.} 
\label{Fig6}
\end{figure}

Figure \ref{Fig6} next displays the spectra approximated from the convolution product of the Fourier transforms of $N_e(t)$ and of ${\vec E}(t)$ for the same relative phases, $\phi = 0$ or $\pi/2$, selected in Fig. \ref{Fig4}. For conciseness, only the total THz spectra given by $|\widehat{\partial_t J}|$ are presented. These are computed from Eq. (\ref{13}) using (\ref{12}) calculated on the exact ionization instants $t_n^{\rm num}$ - thus corresponding to the general expression (\ref{14}) - and they are compared with their approximation $|{\vec E}_{\rm rad}|$ discarding the pulse envelope [Eq. (\ref{15})] and utilizing either $t_n^{\rm num}$ or the analytically-computed instants $t_n^{\rm an}$. We report an excellent agreement between the different curves. Note, however, that for the CP-S case, Eq. (\ref{14b}), including the traces of ionization inside the oscillating components, has been preferred to Eq. (\ref{15}) for reasons detailed below. We also observe that summing up over discrete ionization instants can introduce artifacts that bridge the different spectral humps (see also \cite{Cabrera2015}). Such discrepancies, nonetheless, rather occur at the noise level, as can be seen in Fig. \ref{Fig6}(d).\\

The previous agreement enables us to expand Eq. (\ref{15}) using the ionization times (\ref{11a})-(\ref{11d}) according to the pump polarization state of interest. 

In the LP-P cases, we can extract the resonant modes by introducing $\beta_n \equiv n \omega_2 \pi/\omega_1 + \phi$ where $\omega_2 = \delta \omega + \omega_1$, so that the ionization instants Eqs. (\ref{11a}) and (\ref{11b}) become
\begin{align}
    \label{16a}
    \mbox{LP-P}: & \quad \omega_1 t_n \approx n \pi - \sqrt{\frac{r}{1-r}}(1+\frac{\delta \omega}{\omega_1}) \sin{(\frac{\delta \omega}{\omega_1} n \pi + \phi)}, \\ 
    \mbox{LP-O}: & \quad \omega_1 t_n \approx n \pi - \frac{1}{2} \frac{r}{1-r} (1+\frac{\delta \omega}{\omega_1}) \sin{(2\frac{\delta \omega}{\omega_1} n \pi + 2\phi)}.
    \label{16b}
\end{align}
We moreover observe that all contributions in $\cos{\beta_n}$ and $\sin{\beta_n}$ of the vectorial components in Eq. (\ref{15}) can easily be factorized as a generic vector $(-1)^n \, {\vec G}\left(\cos{(\frac{\delta \omega}{\omega_1} n \pi + \phi)},\sin{(\frac{\delta \omega}{\omega_1} n \pi + \phi)}\right)$. For technical convenience we assume density steps and ionization durations being equal for each $n$. By taking the limit of small $r$ in the exponential of Eq. (\ref{15}) together with $\kappa \rightarrow 0$, the radiated spectra straightforwardly follow from the sums 
\begin{align}
    \label{16c}
    \mbox{LP-P}: & \quad \sum_{n=1}^N \mbox{e}^{i n \pi(\frac{\omega}{\omega_1} - 1)}\left( 1 - i \frac{\omega}{\omega_1} \sqrt{\frac{r}{1-r}}(1+\frac{\delta \omega}{\omega_1}) \sin{(\frac{\delta \omega}{\omega_1} n \pi + \phi)} \right){\vec G}_{\rm LP-P}, \\
    \mbox{LP-O}: & \quad \sum_{n=1}^N \mbox{e}^{i n \pi(\frac{\omega}{\omega_1} - 1)}\left( 1 - i \frac{\omega}{\omega_1}\frac{r}{1-r}(1+\frac{\delta \omega}{\omega_1}) \sin{(2\frac{\delta \omega}{\omega_1} n \pi + 2\phi)} \right){\vec G}_{\rm LP-O}.
    \label{16d}
\end{align}
Therefrom, the Brunel spectrum proceeds from simple geometric series providing after summation:
\begin{equation}
\label{17a}
|{\vec E}_{\rm rad}|(\omega) \propto |\sum_{\pm} \sum_{j=1, 2} \mbox{e}^{\pm i j \phi} \frac{\sin{(\frac{\pi}{2} N b_{j}^{\pm})}}{\sin{(\frac{\pi}{2} b_{j}^{\pm})}}|,
\end{equation}
where $b_{j}^{\pm} = \omega/\omega_1 - 1 \pm j \delta \omega/\omega_1$ and $N$ denotes the total number of ionization events. From the above expression it is clear that resonant modes are excited at the zeroes of $\sin{(\pi b_{j}^{\pm}/2)}$, i.e., when $b_{j}^{\pm} = 2k,\,k \in \Z$ or equivalently $\omega = (2k+1) \omega_1 \pm j \delta \omega$. With $k=0$ one recovers the visible radiation mode $\omega = 2\omega_1 - \omega_2$ from $b_1$ and the mid-IR radiation mode $\omega = 2\omega_2 - \omega_1$ from $b_2$. Both secondary radiations belong to the Brunel spectrum triggered by LP-P pulses. Interestingly, the LP-O radiated spectrum: 
\begin{equation}
     \label{18b}
    {\vec E}_{\rm rad}(\omega) \propto \sum_n \delta N_n \mbox{e}^{i \omega t_n}
    \left(\begin{array}{c}
(-1)^n  \frac{\omega}{\omega^2 - \omega_1^2}  \\
   \sqrt{\frac{r}{1-r}} \frac{1}{\omega^2-\omega_2^2} (\omega \cos{\beta_n} - i \omega_2 \sin{\beta_n})
\end{array}\right),
\end{equation}
has an $x$-component that can only convey the mid-IR radiation while the $y$ component contains both radiated modes, in agreement with Fig. \ref{Fig4}(b). Note that only the $y$-component contributes to the THz spectrum $\omega \rightarrow 0$.

By using similar arguments, we deduce that the circularly polarized pulses supply a radiated field decomposing over the generic sums:
\begin{equation}
    \label{18c}
    \sum_n \delta N_n \mbox{e}^{i \omega t_n} \cos{(\frac{2 n \pi - \phi}{\omega_2/\omega_1 - \epsilon}}),  \sum_n \delta N_n \mbox{e}^{i \omega t_n} \sin{(\frac{2 n \pi - \phi}{\omega_2/\omega_1 - \epsilon}}),
\end{equation}
which, again for equal density steps and ionization durations, contribute to the Brunel spectrum by numerous interference patterns modeled as 
\begin{equation}
    \label{18d}
    |\vec{E}_{\rm rad}|(\omega) \propto | \sum_\pm \sum_i E_{0,i} \frac{\mbox{e}^{\pm i \phi \gamma_\pm^i}}{\omega \pm \omega_i} \frac{\sin{(N \pi \gamma_\pm^i)}}{\sin{(\pi \gamma_\pm^i)}}|, \,\,\,\, \gamma_\pm^i \equiv \frac{\omega \pm \omega_i}{\omega_2 - \epsilon \omega_1}.
\end{equation}
From Eq. (\ref{18d}) the excitation of new wavelengths can be expected at the maximum of the interference functions, that is to say when $\omega$ satisfies $\omega \pm \omega_i = k (\omega_2 - \epsilon \omega_1)$, where $k \in \Z$. In CP-S configurations ($\epsilon = 1$), the previous resonance condition allows the occurrence of visible and mid-IR radiations at $\omega = 2 \omega_i - \omega_j$ ($i\neq j = 1,2$). It can easily be checked that such resonance conditions cannot be achieved in CP-C configurations ($\epsilon = -1$), which only permits the excitation of higher-order frequency combinations $2\omega_1+\omega_2$ and $2\omega_2+\omega_1$. The faster decrease of the CP-S spectra at large $\omega$ is worth being noticed. Whereas the radiated field $|{\vec E}_{\rm rad}|$ decreases like $1/\omega$ for linearly polarized pulses, its decay is indeed faster than $\sim |\sin(N \pi \gamma_\pm ^1)|/\omega^2$ out of two consecutive excited resonances in the CP-S case. This decrease, however, is overestimated by Eq. (\ref{15}) for this polarization state (not shown), which needs to include the contributions linked to the ionization duration $\tau_n$ - thus to use Eq. (\ref{14b}) instead - to provide a good agreement with the LC results.

\section{About the classical $\omega-2\omega$ scheme}
\label{sec4}

\begin{figure}[ht!]
\centering \includegraphics[width=\columnwidth]{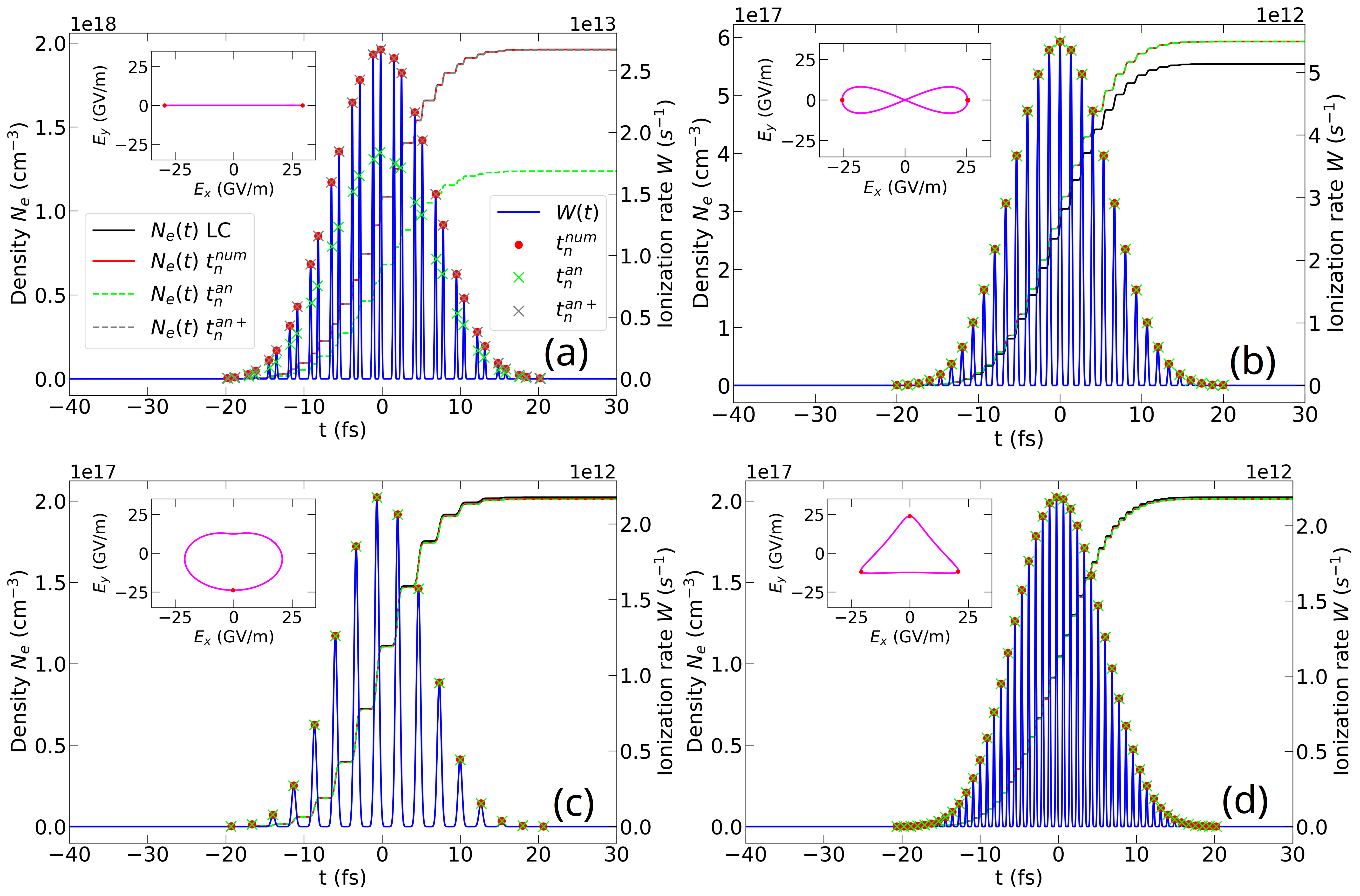} 
\caption{(a-d) Density profiles (black solid curves, left axis), ionization rates (blue solid curves, right axis) and field polarization patterns in the $(E_x,E_y)$ plane (insets) of 35-fs, two-color Gaussian pulses for the four baseline polarization geometries (see text for the laser-gas parameters) when $\omega_2 = 2\omega_1$. The two-color relative phase is $\phi = \pi/2$. The (exact or analytical) ionization times $t_n$ are plotted within the time window from -21 fs to +21 fs. The green crosses mark the time instants evaluated by Eqs. (\ref{11a})-(\ref{11d}); the grey ones refer to the corrected instants Eq. (\ref{24}) in the LP-P case. The Lissajou figures displayed as insets all run over the same optical period: $0-2.7$ fs.}
\label{Fig7}
\end{figure}

\begin{figure}[ht!]
\centering \includegraphics[width=\columnwidth]{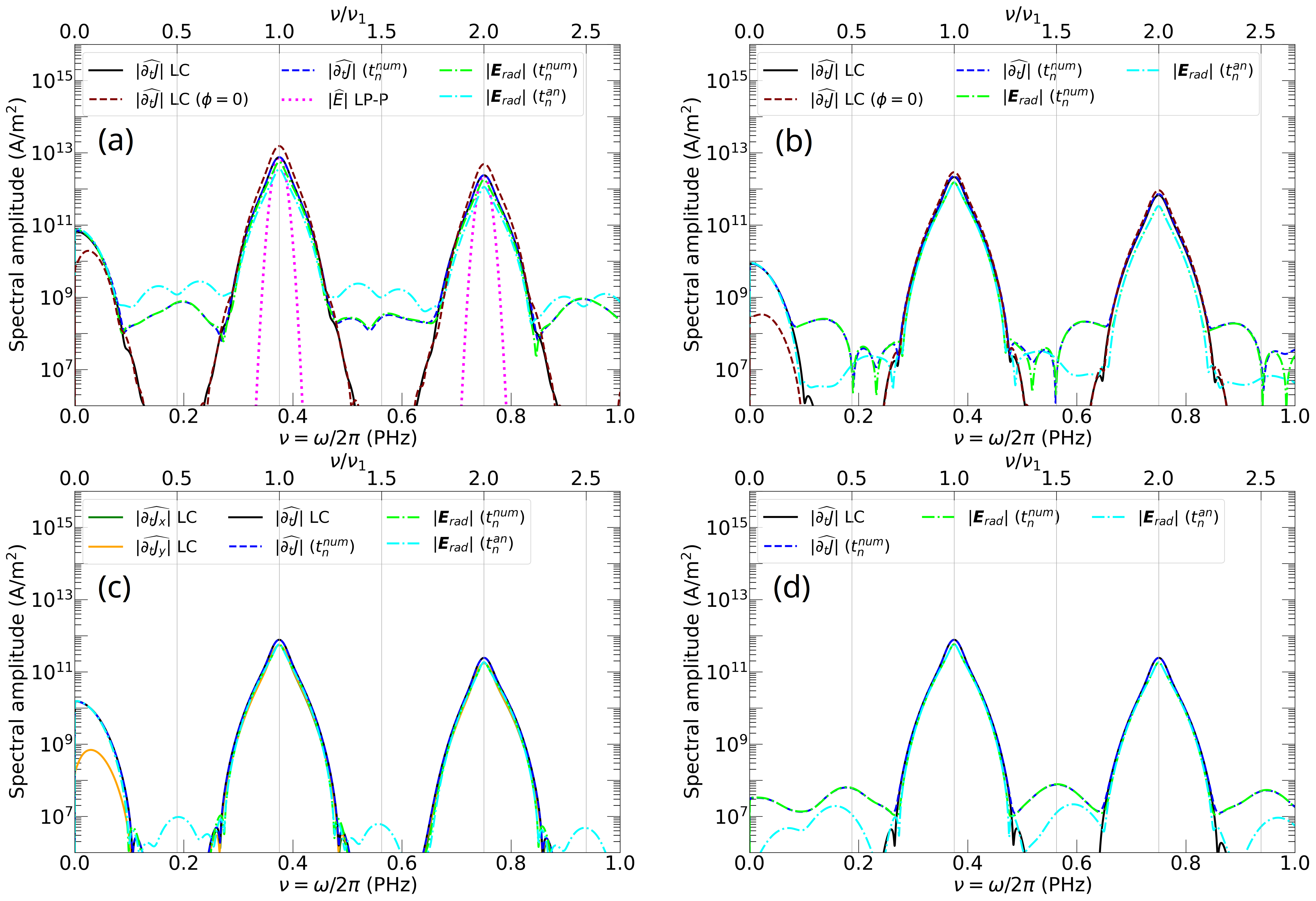}
\caption{Brunel spectra numerically-computed from the LC model for the polarization configurations (a) LP-P, (b) LP-O, (c) CP-S, and (d) CP-C as functions of $\nu = \omega/2 \pi$ when $\omega_2 = 2\omega_1$. The spectral amplitudes of the current components $\widehat{\partial_t J_x}$ and $\widehat{\partial_t J_y}$ (for CP-S pulses only), of the total induced current density $\widehat{\partial_t J}$ (see legend) and of the input pump with maximum lined up on max$\widehat{|\partial_t J}|$ for $\phi = \pi/2$ (pink dotted lines, LP-P case only) are shown. Brown dashed curves illustrate $|\widehat{\partial_t J}|$ for a zero relative phase. The pump spectra plotted from Eq. (\ref{12}) have been regularized by choosing $\kappa = 5 \times 10^{-2}$.} 
\label{Fig8}
\end{figure}

A topic of regular interest is the mixing of laser harmonics, multiple of the fundamental frequency $\omega_1$ \cite{Gonzalez2015}, which leads to THz generation ($\omega \rightarrow 0$) in the classical $\omega-2\omega$ scheme. Repeating the previous analysis results in Figs. \ref{Fig7} and \ref{Fig8} for the density profiles and the Brunel spectra either directly computed from the LC model or approximated from Eqs. (\ref{9b}) and (\ref{15}) using the ionization instants $t_n^{\rm num}$ or $t_n^{\rm an}$ [Eqs. (\ref{11a})-(\ref{11d})]. In this context, since the rapid field variations are $2\pi/\omega_1$-periodic, an interesting alternative to the former expansion of $N_e(t)$ in discrete density steps is to compute the Fourier series of the ionization rate, which, with two colors, decompose over all harmonics of the fundamental $\omega_1$ field. With $\phi \neq 0$ the two-color waveform is not an even function of time, so that the general Fourier series applies:
\begin{align}
    \label{18}
    W(t) & = \frac{w_0}{2} + \sum_{l=1}^{+\infty} \left( w_l^c \cos{(l\omega_1t)} + w_l^s \sin{(l\omega_1t)} \right), \\
    w_l^c & = \frac{\omega_1}{\pi} \int_{-\pi/\omega_1}^{+\pi/\omega_1} W(t) \cos{(l\omega_1t)} dt, \\
    w_l^s & =  \frac{\omega_1}{\pi} \int_{-\pi/\omega_1}^{+\pi/\omega_1} W(t) \sin{(l\omega_1t)} dt.
\end{align}
By using Eq. (\ref{8}) in which the Gaussian envelopes of ${\vec E}(t)$ are discarded, straightforward calculations lead to 
\begin{equation}
     \label{20}
    W(t) \approx \frac{w_0}{2} + \frac{\omega_1}{2\sqrt{\pi}} \sum_n \tau_n W(t_n) \sum_{l=1}^{+\infty} \mbox{e}^{- \frac{l^2 \omega_1^2 \tau_n^2}{4}} \cos{(l\omega_1(t-t_n))}
\end{equation}
under the conditions $\omega_1 t_n \neq \pm \pi$ and $\omega_1 \tau_n \ll 1$. The electron density then follows as
\begin{equation}
\label{21}
N_e(t) = N_a (1 - \mbox{e}^{-\frac{N_e^{\rm unsat}(t)}{N_a}}),
\end{equation}
where
\begin{equation}
    \label{22}
    N_e^{\rm unsat}(t) \approx \frac{w_0 N_a}{2} t + \sum_n \sum_{l = 1}^{+\infty} \frac{\mbox{e}^{-l^2 \omega_1^2 \tau_n^2/4}}{\pi l} \delta N_n \sin{(l \omega_1 (t-t_n))}.
\end{equation}
The previous expression of $N_e(t)$ can be used to immediately infer from its product with the pump field expressions (\ref{5})-(\ref{7}) that i -- all harmonics (not only the even ones) are excited in the density spectrum and ii -- the resonant $k$th harmonic is given when $\omega_2 = j\omega_1$ by the integers $l_1 \pm 1 = k,\,l_2 \pm j=k,\,j\neq 1$, that is to say, with $\omega_2 = 2\omega_1$, $l_1 = 2,4$ and $l_2 = 1,5$ for third-harmonic generation, and $l_1 = 1$, $l_2 = 2$ for THz, $0$th harmonic generation. When assuming $r\ll1$ and employing Eqs. (\ref{11a})-(\ref{11d}) for this latter case, Eq. (\ref{22}) multiplied by Eqs. (\ref{6}) and (\ref{7}) restores the expected behaviors: $E_{\rm rad} \propto \frac{3}{2} \sqrt{r/(1-r)} \sin{\phi}$ for the LP-P case, $E_{\rm rad} \propto \frac{1}{2} \sqrt{r/(1-r)} \sin{\phi}$ for the LP-O configuration, together with the vectorial components of the (linearly polarized) THz field: 
\begin{align}
\label{23}
{\vec E}_{\rm rad} \propto
\left(\begin{array}{c}
 \sin{\phi} \\
 \cos{\phi}
\end{array}\right),\left(\begin{array}{c}
 -\sin{(2n\pi/3 - \phi/3)} \\
 \cos{(2n\pi/3 - \phi/3)}
\end{array}\right),
\end{align}
generated by the CP-S and CP-C pulses, respectively \cite{Tailliez2020}. Note that higher harmonics have their respective amplitude decreasing at leading order like $1/l$, in agreement with Brunel's theory for a single-color pulse \cite{Brunel1990}.

Again utilizing Eqs. (\ref{11a})-(\ref{11d}) with $\omega_2 = 2\omega_1$ furthermore enables us to recover easily the main density dynamics and spectral amplitudes of two-colors-driven THz pulses in the $\omega-2\omega$ scheme from both Eqs. (\ref{13}) and (\ref{15}). Their approximations generally agree with the LC results shown in Figs. \ref{Fig7} and \ref{Fig8} for a $0.8\,\mu$m fundamental pulse and a $0.4\,\mu$m second harmonic ($\omega_2 = 2 \omega_1)$. In Fig. \ref{Fig7}(a), however, we report a degraded growth in the electron density in the LP-P configuration (see green dashed curve) when approaching the latter with the O($r$)-truncated approximation $t_n^{\rm an}$ defined by Eq. (\ref{11a}). Figure \ref{Fig7}(a) reveals that among the two ionization events per cycle expected in this case, only one corresponds to the maximum of $W(t)$ whereas the other, slightly shifted in time, does not contribute when $t_n$ is provided by Eq. (\ref{11a}). This explains the deviation in the density accumulated in $N_e(t)$ when this rough approximation is used. For the optimized relative phase value $\phi = \pi/2$, it is possible, nonetheless, to find exact roots to $\partial_t |{\vec E}| = 0$ from Eq. (\ref{6}) in the LP-P case expressing as 
\begin{equation}
    \label{24}
    \omega_1 t_n^{an+} = n \pi - (-1)^n \arcsin{\left( \frac{\sqrt{1-r}}{8 \sqrt{r}}(\sqrt{\frac{32 r}{1-r}+1} -1) \right)}.
\end{equation}

\noindent Employing instead the above corrected instant restores density curves and ionization instants agreeing much better with the LC results (compare green dashed and grey solid curves and crosses in Fig. \ref{Fig7}(a); the latter curve superimposes with the LC data). Such discrepancies, however, strongly diminish in the LP-O case [Fig. \ref{Fig7}(b)], which we attribute to the ordering of $t_n^{\rm an}$ in $O(r^2)$ [see Eq. (\ref{11b})] providing a better approximation than the instants limited to order $O(r)$ for LP-P pulses. Again, we observe an excellent agreement in the circularly polarized configurations between the density profiles evaluated with the exact $t_n$ and their analytical version [see Figs. \ref{Fig7}(c) and \ref{Fig7}(d)].

About the spectra displayed in Fig. (\ref{Fig8}), the sensitivity of linearly polarized pulses to the relative phase appears amplified in the $\omega-2\omega$ scheme [Figs. \ref{Fig8}(a) and \ref{Fig8}(b)], compared to the previous setup of non-harmonic frequency mixing. The spectral amplitude in the LP-O case diminishes at small frequencies by a factor $\lesssim 1/10$ compared with a LP-P pulse. In the CP-S configuration [Fig. \ref{Fig8}(c)] changes only manifest in the THz region by the reverted contributions of the $x-$ and $y-$components of the current components that are aligned with the polarization angle of the generated THz pulse along the vector $(\sin{\phi},\cos{\phi})^T$ \cite{Tailliez2020}. Finally, although the approximation $|{\vec E}_{\rm rad}|$ based on the instants $t_n^{\rm an}$ can become slightly degraded compared with the LC spectra or their counterparts employing the exact ionization instants $t_n^{\rm num}$, such discrepancies only occur at low spectral amplitudes close to the noise level [see, e.g., Fig. \ref{Fig8}(d)].

\section{Conclusion}
\label{sec5}
In summary, we have proposed a simple approach to describe the ionization dynamics induced by a two-color ionizing pulse when the latter interacts with an atom gas. For an electron density growing steplike the key ingredient is the accurate evaluation of the ionization instants. The radiation spectra then follow from the convolution product between the density spectrum and the pump field spectrum -- demonstrated to be valid by full quantum TDSE calculations. Our analysis has shown that an approximation relying on a Taylor expansion of the laser electric field truncated at second order is sufficient to provide a very good agreement of the electron densities and Brunel spectra deduced from the exact ionization instants. It appears that the approximation of these instants, when one ignores the variations of the pulse envelope and applies a first-order development in the amplitude ratio of the two colors, generally reproduces rather well these quantities numerically computed from the local current model, except in the $\omega-2\omega$ scheme for linearly polarized parallel fundamental and second harmonic pulses. We have shown that the knowledge of such approximated ionization instants is, nevertheless, useful to describe and explain the emergence of secondary radiations. Finally, it is worth emphasizing that the two incommensurable frequency mixing presented in this work leads to an ultrabroad spectrum spanning from the THz to the visible range -- and even beyond -- with peaks of comparable magnitude to those of the common $\omega-2\omega$ two-color mixing, which has driven so far an important literature in the field of terahertz science.


\begin{thebibliography}{00}

\bibitem{Brunel1990}
  F. Brunel, {\it Harmonic generation due to plasma effects in a gas undergoing multiphoton ionization in the high-intensity limit}, J. Opt. Soc. Am. B {\bf 7}, 521- 526 (1990).

\bibitem{Berge2007}
    L. Berg{\'e}, S. Skupin, R. Nuter, J. Kasparian, and J.-P. Wolf, {\it Ultrashort filaments of light in weakly ionized, optically transparent media}, Rep. Prog. Phys. {\bf 70} 1633–1713 (2007).

\bibitem{Kim2008}
  K. Y. Kim, A. J. Taylor, J. H. Glownia, and G. Rodriguez, {\it Coherent control of terahertz supercontinuum generation in ultrafast laser–gas interactions}, Nat. Photonics {\bf 2}, 605-609 (2008).

\bibitem{Wang2009}
    T.-J. Wang, Y. Chen, C. Marceau, F. Th{\'e}berge, M. Ch{\^a}teauneuf, J. Dubois, and S. L. Chin, {\it High energy terahertz emission from two-color laser-induced filamentation in air with pump pulse duration control}, Appl. Phys. Lett. {\bf 95}, 131108 (2009).
    
\bibitem{Liu2010}
    J. Liu, J. Dai, S. L. Chin, and X.-C. Zhang, {\it Broadband terahertz wave remote sensing using coherent manipulation of fluorescence from asymmetrically ionized gases}, Nat. Photon. {\it 4}, 627-631 (2010).

\bibitem{Babushkin2010}
  I. Babushkin, W. Kuehn, C. K{\"o}hler, S. Skupin, L. Berg{\'e}, K. Reimann, M. Woerner, J. Herrmann, and T. Elsaesser, {\it Ultrafast Spatiotemporal Dynamics of Terahertz Generation by Ionizing Two-Color Femtosecond Pulses in Gases}, Phys. Rev. Lett. {\bf 105}, 053903 (2010).

\bibitem{Babushkin2011}
   I. Babushkin, S. Skupin, A. Husakou, C. K{\"o}hler, E. Cabrera-Granado, L. Berg{\'e}, and J. Herrmann, {\it Tailoring terahertz radiation by controling tunnel photoionization events in gases}, New J. Phys. {\bf 13}, 123029 (2011).

\bibitem{Gonzalez2015}
    P. Gonz{\'a}lez de Alaiza Mart{\'i}nez, I. Babushkin, L. Berg{\'e}, S. Skupin, E. Cabrera-Granado, C. K{\''o}hler, U. Morgner, A. Husakou, and J. Herrmann, {\it Boosting Terahertz Generation in Laser-Field Ionized Gases Using a Sawtooth Wave Shape}, Phys. Rev. Lett. {\bf 114}, 183901 (2015).

\bibitem{Clerici2013}
    M. Clerici, M. Peccianti, B. E. Schmidt, L. Caspani, M. Shalaby, M. Gigu{\'e}re, A. Lotti, A. Couairon, F. L{\'e}gar{\'e}, T. Ozaki, D. Faccio, and R. Morandotti, {\it Wavelength scaling of terahertz generation by gas ionization}, Phys. Rev. Lett. {\bf 110}, 253901 (2013).

\bibitem{Nguyen2017}
    A. Nguyen, P. Gonz{\'a}lez de Alaiza Mart{\'i}nez, J. D{\'e}chard, I. Thiele, I. Babushkin, S. Skupin, and L. Berg{\'e}, {\it Spectral dynamics of THz pulses generated by two-color laser filaments in air: the role of Kerr nonlinearities and pump wavelength}, Opt. Express {\bf 25}, 4720-4740 (2017).

\bibitem{Nguyen2018_2}
    A. Nguyen, P. Gonz{\'a}lez de Alaiza Mart{\'i}nez, I. Thiele, S. Skupin, and L. Berg{\'e}, {\it Broadband terahertz radiation from two-color mid- and far-infrared laser filaments in air}, Phys. Rev. A {\bf 97}, 063839 (2018).

\bibitem{Damico2007}
    C. D’Amico, A. Houard, M. Franco, B. Prade, A. Mysyrowicz, A. Couairon, and V. T. Tikhonchuk, {\it Conical forward THz emission from femtosecond-laser-beam filamentation in air}, Phys. Rev. Lett. {\bf 98}, 235002 (2007).

\bibitem{Andreeva2016}
     V. A. Andreeva, O. G. Kosareva, N. A. Panov, D. E. Shipilo, P. M. Solyankin, M. N. Esaulkov, P. Gonz{\'a}lez de Alaiza Mart{\'i}nez, A. P. Shkurinov, V. A. Makarov, L. Berg{\'e}, and S. L. Chin, {\it Ultrabroad Terahertz Spectrum Generation from an Air-Based Filament Plasma}, Phys. Rev. Lett. {\bf 115}, 063902 (2016).

\bibitem{Nikolaeva2024}
    I. A. Nikolaeva, N. R. Vrublevskaya, G. E. Rizaev, D. V. Pushkarev, D. V. Mokrousova, D. E. Shipilo, N. A. Panov, L. V. Seleznev, A. A. Ionin, O. G. Kosareva, and A. B. Savel'ev, {\it Terahertz ring beam independent on x–2x phase offset in the course of two-color femtosecond filamentation}, Appl. Phys. Lett. {\bf 124}, 051105 (2024).

\bibitem{Vaicaitis2019}
    V. Vai\v{c}aitis, O. Balachninait{\.e}, U. Morgner, and I. Babushkin, {\it Terahertz radiation generation by three-color laser pulses in air filament}, J. Appl. Phys. {\bf 125}, 173103 (2019).

\bibitem{Babushkin2022}
    I. Babushkin, A. J. Gal{\'a}n, J. R. C. de Andrade, A. Husakou, F. Morales, M. Kretschmar, T. Nagy, V. Vai\v{c}aitis, L. Shi, D. Zuber, et al., {\it All-optical attoclock for imaging tunnelling wavepackets}, Nat. Physics {\bf 18}, 417–422 (2022).

\bibitem{Kostin2019}
    V. A. Kostin and N. V. Vvedenskii, {\it Mutual enhancement of Brunel harmonics}, JETP Lett. {\bf 110}, 457-463 (2019).

\bibitem{Kostin2020}
    V. A. Kostin, I. D. Laryushin, and N. V. Vvedenskii, {\it Generation of Terahertz Radiation by Multicolor Ionizing Pulses}, JETP Lett., {\bf 112}, 77–83 (2020).

\bibitem{Vvedenskii2014}
    N. V. Vvedenskii, A. I. Korytin, V. A. Kostin, A. A. Murzanev, A. A. Silaev, and A. N. Stepanov1, {\it Two-Color Laser-Plasma Generation of Terahertz Radiation Using a Frequency-Tunable Half Harmonic of a Femtosecond Pulse} {\it 112}, 055004 (2014).

\bibitem{Kostin2016}
    V. A. Kostin, I. D. Laryushin, A. A. Silaev, and N. V. Vvedenskii, {\it Ionization-induced multiwave mixing: Terahertz generation with two-color laser pulses of various frequency ratios}, Phys. Rev. Lett. {\bf 117}, 035003 (2016).

\bibitem{Vaicaitis2018}
    V. Vai\v{c}aitis and V. Tamulien{\.e}, {\it Four-wave mixing in air by bichromatic spectrally broadened femtosecond laser pulses}, J. Opt. Soc. Am. B {\bf 35}, 2091-2095 (2018).

\bibitem{Kim2009}
    K. Y. Kim, {\it Generation of coherent terahertz radiation in ultrafast laser-gas interactions}, Phys. Plasmas {\bf 16}, 056706 (2009).

\bibitem{Wang2017}
    W.-M. Wang, Z.-M. Sheng, Y.-T. Li, Y. Zhang, and J. Zhang, {\it Terahertz emission driven by two-color laser pulses at various frequency ratios}, Phys. Rev. A {\bf 96}, 023844 (2017).
    
\bibitem{Zhang2017}
    L.-L. Zhang, W.-M. Wang, T. Wu, R. Zhang, S.-J. Zhang, C.-L. Zhang, Y. Zhang, Z.-M. Sheng, and X.-C. Zhang, {\it Observation of Terahertz Radiation via the Two-Color Laser Scheme with Uncommon Frequency Ratios}, Phys. Rev. Lett. {\bf 119}, 235001 (2017).

\bibitem{Agrawal2012}
    G. Agrawal, {\it Nonlinear Fiber Optics}, Academic Press, New York (2012).

\bibitem{Theberge2006}
    F. Th{\'e}berge, N. Ak{\"o}zbek, W. Liu, A. Becker, and S.-L. Chin, {\it Tunable Ultrashort Laser Pulses Generated through Filamentation in Gases}, Phys. Rev. Lett. {\bf 97}, 023904 (2006).  

\bibitem{Theberge2010}
    F. Th{\'e}berge, M. Ch{\^a}teauneuf, G. Roy, P. Mathieu, and J. Dubois, {\it Generation of tunable and broadband far-infrared laser pulses during two-color filamentation}, Phys. Rev. A {\bf 81}, 033821 (2010).
    
\bibitem{Li2012}
    M. Li, W. Li, Y. Shi, P. Lu, H. Pan, and H. Zeng, {\it Verification of the physical mechanism of THz generation by
    dual-color ultrashort laser pulses}, Appl. Phys. Lett. {\bf 101}, 161104 (2012).

\bibitem{Wang2022}
    S. Wang, C. Lu, Z. Fan, A. Houard, V. Tikhonchuk, A. Mysyrowicz, S. Zhuang, V. A. Kostin, and Y. Liu, {\it Coherently controlled ionization of gases by three-color femtosecond laser pulses}, Phys. Rev. Lett. {\bf 105}, 023529 (2022).

\bibitem{Ammosov1986}
    M. V. Ammosov and N. B. Delone and V. P. Kra\v{i}nov, {\it Tunnel ionization of complex atoms and of atomic ions in an alternating electromagnetic field}, Sov. Phys. JETP {\bf 64}, 1191-1194 (1986).

\bibitem{Nguyen2018}
    A. Nguyen, P. Gonz{\'a}lez de Alaiza Mart{\'i}nez, I. Thiele, S. Skupin and L. Bergé, {\it THz field engineering in two-color femtosecond filaments using chirped and delayed laser pulses}, New J. Phys. {\bf 20}, 033026 (2018).

\bibitem{Jefimemko1989}
    O. D. Jefimenko, {\it Electricity and Magnetism: An Introduction to the Theory of Electric and Magnetic Fields}, (Electret Scientific Company, 1989).

\bibitem{Rae1992}
    S. C. Rae and K. Burnett, {\it Detailed simulations of plasma-induced spectral blueshifting}, Phys. Rev. A {\bf 46}, 1084-1090 (1992).

\bibitem{Landau1965}
    L. M. Landau and E. M. Lifschitz {\it Quantum Mechanics} 2nd Ed. (New York - Pergamon) p 276 (1965).

\bibitem{Gonzalez2014}
    P. Gonz{\'a}lez de Alaiza Mart{\'i}nez and L. Berg{\'e}, {\it Influence of multiple ionization in laser filamentation}, J. Phys. B: At. Mol. Opt. Phys. {\bf 47}, 204017 (2014).

\bibitem{Meng2016}
    C. Meng, W. Chen, X. Wang, Z. L{\"u}, Y. Huang, J. Liu, D. Zhang, Z. Zhao, and J. Yuan, {\it Enhancement of terahertz radiation by using circularly polarized two-color laser fields}, Appl. Phys. Lett. {\bf 109}, 131105 (2016).

\bibitem{Tailliez2020}
    C Tailliez, A Stathopulos, S Skupin, D. Buo\v{z}ius, I Babushkin,
    V. Vai\v{c}aitis and L Berg{\'e}, {\it Terahertz pulse generation by two-color laser fields with circular polarization}, New J. Phys. {\bf 22}, 103038 (2020).

\bibitem{Stathopulos2024}
A. Stathopulos, S. Skupin, B. Zhou, P. U. Jepsen, and L. Berg{\'e}, {\it Waveshape of terahertz radiation produced by two-color laser-induced air plasmas} Phys. Rev. Res. {\bf 4}, 043274 (2024).

\bibitem{Catoire2016}
  F. Catoire, A. Ferré, O. Hort, A. Dubrouil, L. Quintard, D. Descamps, S. Petit, F. Burgy, E. Mével,Y. Mairesse, and E. Constant, {\it Complex structure of spatially resolved high-order-harmonic spectra}, Phys. Rev. A {\bf 94}, 063401 (2016).

\bibitem{Vabek2022}
    J. V{\'a}bek, H. Bachau, and F. Catoire, {\it Ionization dynamics and gauge invariance}, Phys. Rev. A {\bf 106}, 053115 (2022).

\bibitem{Finke2022}
  O. Finke, J. Vábek, M. Nevrkla, N. Bobrova, O. Hort, L. Jurkovičová, M. Albrecht, A. Jančárek, F. Catoire, S. Skupin, and  J. Nejdl, {\it Phase-matched high-order harmonic generation in pre-ionized noble gases}, Sci. Rep. {\bf 12}, 7715 (2022).

\bibitem{Thiele2017}
    I. Thiele, P. Gonz{\'a}lez de Alaiza Mart{\'i}nez, R. Nuter, A. Nguyen, L. Berg{\'e} and S.Skupin, {\it Broadband terahertz emission from two-color femtosecond-laser-induced microplasmas} Phys. Rev. A {\bf 96}, 053814 (2017).

\bibitem{Cabrera2015}
E. Cabrera-Granado, Y. Chen, I. Babushkin, L. Berg{\'e}, and Stefan Skupin,  {\it Spectral self-action of THz emission from ionizing two-color laser
pulses in gases}, New J. Phys. {\bf 17}, 023060 (2015).

\end{thebibliography}
\end{document}